\documentclass{aa}

\usepackage[utf8]{inputenc}
\usepackage{tabularx}
\usepackage{multirow}
\usepackage{url}
\usepackage{amsmath}    
\usepackage{amssymb}    
\usepackage{graphicx}
\usepackage{color}
\usepackage{natbib}
\usepackage{hyperref}
\usepackage{times,txfonts}
\usepackage{bm}
\usepackage{ae,aecompl}

\bibpunct{(}{)}{;}{a}{}{,} 

\usepackage{listingsutf8}

\newcommand{\codeil}[1]{\lstinline!#1!}

\newcommand{\Cfl}{\hat{c}}
\newcommand{\hf}{\frac{1}{2}}

\newcommand{\Bhat}{\vec{\hat{B}}}
\newcommand{\Ehat}{\vec{\hat{E}}}
\newcommand{\Jhat}{\vec{\hat{J}}}
\newcommand{\uhat}{\vec{\hat{u}}}
\newcommand{\vhat}{\vec{\hat{v}}}
\newcommand{\xhat}{\vec{\hat{x}}}
\newcommand{\that}{\hat{t}}

\newcommand{\gammam}{\langle \gamma \rangle}

\newcommand{\f}[1]{\ensuremath{f_{\mathrm{{#1}}}}}

\newcommand{\derfrac}[2]{\frac{\ud #1}{\ud #2}} 
\newcommand{\parfrac}[2]{\frac{\partial #1}{\partial #2}} 

\newcommand{\runko}{\textsc{runko}}
\newcommand{\corgi}{\textsc{corgi}}
\newcommand{\tristan}{\textsc{Tristan}}
\newcommand{\tristanmp}{\textsc{Tristan-MP}}
\newcommand{\pybind}{\textsc{Pybind11}}

\newcommand{\numpy}{\textsc{Numpy}}
\newcommand{\scipy}{\textsc{Scipy}}
\newcommand{\matplotlib}{\textsc{Matplotlib}}
\newcommand{\python}{\textsc{Python}}

\newcommand{\be}{\begin{equation}}
\newcommand{\ee}{\end{equation}}
\newcommand{\Ten}[2]{\ensuremath{#1 \times 10^{#2}} }
\newcommand{\unitspace}{\,}

\newcommand{\cm}{\ensuremath{\unitspace \mathrm{cm}}}

\newcommand{\Gauss}{\ensuremath{\unitspace \mathrm{G}}}

\newcommand{\nvec}[1]{\ensuremath{\hat{\bm{\mathit{#1}}}}}
\renewcommand{\vec}[1]{\ensuremath{\bm{\mathit{#1}}}}

\newcommand{\ud}{\mathrm{d}}

\newcommand{\cpp}{C\texttt{++}}
\newcommand{\cppv}[1]{C\texttt{++}$#1$}

\begin{document}

\title{\textsc{Runko}: Modern multiphysics toolbox for plasma simulations}

\titlerunning{\runko: Modern plasma simulation framework}

\author{J.\,N\"attil\"a\inst{1,2}}

\institute{
    $^1$ Physics Department and Columbia Astrophysics Laboratory, Columbia University, 538 West 120th Street, New York, NY 10027, USA.\\
    $^2$ Center for Computational Astrophysics, Flatiron Institute, 162 Fifth Avenue, New York, NY 10010, USA. \\
    \email{jnattila@flatironinstitute.org}
}

\date{Received XXX / Accepted XXX}

\abstract{
    \runko\ is a new open-source plasma simulation framework implemented in \cpp\ and \python.
    It is designed to function as an easy-to-extend general toolbox for simulating astrophysical plasmas with different theoretical and numerical models.
    Computationally intensive low-level kernels are written in modern \cpp\ taking advantage of polymorphic classes, multiple inheritance, and template metaprogramming.
    High-level functionality is operated with \python\ scripts.
    The hybrid program design ensures good code performance together with ease of use. 
    The framework has a modular object-oriented design that allows the user to easily add new numerical algorithms to the system.
    The code can be run on various computing platforms ranging from laptops (shared-memory systems) to massively parallel supercomputer architectures (distributed-memory systems).
    The framework supports heterogeneous multiphysics simulations in which different physical solvers can be combined and run simultaneously.
    Here we showcase the  framework's relativistic particle-in-cell (PIC) module by presenting  
    (\textit{i}) 1D simulations of relativistic Weibel instability, 
    (\textit{ii}) 2D simulations of relativistic kinetic turbulence in a suddenly stirred magnetically-dominated pair plasma, and
    (\textit{iii}) 3D simulations of collisionless shocks in an unmagnetized medium.
}

\keywords{ Plasmas -- Turbulence -- Methods: numerical }

\maketitle

\section{Introduction}\label{sec:intro}

Ever since the introduction of computers, numerical simulations have been used to study the nonlinear behavior of plasma \citep[see, e.g.,][]{Buneman_1959, Dawson_1962, Dawson_1964}.
In the early days, the research was mainly motivated by studies of basic plasma instabilities and confinement in fusion experiments, but the method of computationally solving the dynamics of charged particles quickly gained popularity also in understanding plasma in space \citep[see, e.g.,][]{Tanaka_1983,Langdon_1988,Buneman_1992}.

An important factor in accelerating the use of plasma simulations is the ever-increasing computational speed and number of processors.
We have recently arrived in the petaflop  ($10^{15}$ floating-point operations per second, FLOPS) supercomputing era, and have started to pave our way toward  exascale computing systems ($10^{18}$ FLOPS). 
This technological surge (and the ingenuity of the researchers themselves) has enabled us to shed light on many longstanding issues in high-energy astrophysics, including 
shocks \citep[e.g.,][]{Frederiksen_2004, spitkovsky2005}, 
reconnection \citep[e.g.,][]{Zenitani_2001, Cerutti_2012, Kagan_2013, sironi2014, Werner_2015}, 
pulsar magnetospheres \citep[e.g.,][]{Philippov_2014, Chen_2014a, Cerutti_2015},
kinetic turbulence \citep[e.g.,][]{Howes_2008, Servidio_2014, Zhdankin_2017a, Comisso_2018, Nattila_2021}, and
nonlinear wave interactions \citep[e.g.,][]{Nattila_2022}.

In order to keep riding this technological wave, successful numerical codes need to be highly scalable and modern in order to take advantage of the newest supercomputing architectures.
These kinds of exascale-ready features include avoidance of global communication patterns and support of heterogeneous computing nodes with varying resources, such as a different number of cores/accelerators or varying amount of memory per core (see also \citealt{Nordlund_2018}, for a discussion). 
The presented \runko\ framework was developed from scratch to support these features.

An additional feature of the plasma simulations is that there are various theoretical frameworks for modeling the dynamics of the system, whereas the underlying numerical algorithms all rely on techniques based on computational particles and multi-dimensional grids.
The fully kinetic description is based on sampling the distribution function with superparticles and representing the electromagnetic fields on a grid, known as the  particle-in-cell (PIC) approach \citep{Hockney_1981, birdsall1985}.%
\footnote{Alternatively, such a system can be modeled by solving the partial differential equations directly on a grid (known as the Vlasov method; \citealt{Cheng1976}).
}
In the  case of a strong (background) magnetic field, it is possible to average over the gyrorotation of the particle's gyro-orbit and use the  gyrokinetic description where, instead of particles, computational ``rings'' are considered \citep[see, e.g.,][]{Howes_2006}.
Alternatively, the kinetic systems can be significantly simplified by considering only the moments of the distribution function;
this macro-scale description is related to the magnetohydrodynamic (MHD) models \citep[see, e.g.,][]{Brandenburg_2003}.
The MHD description itself can be divided into multiple categories with different closure approximations (e.g., ideal, visco-resistive) or the evolution of  multiple fluid species (known as multifluid description).
There also exists a class of hybrid methods that combine both kinetic and magnetohydrodynamic models, for example by treating electrons with fluid description and protons with the kinetic description \citep[see, e.g.,][]{Winske_2003, Valentini_2007, Haggerty_2019}.
From the numerical perspective, all of these methods, in the end, only require the  efficient manipulation of particles and/or physical quantities that are stored on grids.

It is possible to take advantage of this structural similarity of the various formulations with well-designed software abstractions.
This is the second core idea of the \runko\ simulation framework, which  in  essence is just a collection of data structures for storing, manipulating, and communicating particles and grids efficiently.
The framework relies on modern computer science methods and paradigms that provide a sufficient degree of problem generalization to technically enable easy introduction of physics knowledge into the simulations.
Technically this is achieved by polymorphic multiple inheritance as supported by the use of modern \cpp.
The concept of object inheritance also nicely reflects the mathematical way of constructing hierarchical models that build on   the previous one.
Another level of abstraction is obtained by using  template metaprogramming that allows, for example, the  parameterization of  the dimensionality of the system, again closely reflecting the mathematical form of the problem.
Finally, in order to balance the computational efficiency and the user-friendliness of the code, the fast low-level \cpp\ classes are exposed to high-level \python3 scripting language.

The complete framework is provided as open-source software for the community.
It is available from a GitHub repository%
\footnote{%
\url{https://github.com/natj/runko} 
}
and the code design and structure are presented in detail in this article.
Here we also present the PIC code of the framework and describe the implementation of the solvers needed to operate it.
First, the theoretical background of the kinetic modeling of plasma is briefly reviewed in Sect.~\ref{sect:theory}.
The numerical implementation of the framework is then discussed in Sect.~\ref{sect:numerics}.
As an example, the use of the new PIC code is demonstrated on astrophysical plasma problems. 
A discussion of new features and the future directions of the framework are given in Sect.~\ref{sect:discussion}.
Finally, the content is summarized in Sect.~\ref{sect:summary}.

\section{Theory}\label{sect:theory}

\subsection{Kinetic plasma theory}

The kinetic plasma theory we adopt here is based on the special relativistic formulation of the Vlasov--Boltzmann equation.
The spatial coordinate location vector is given by $\vec{x} \equiv (x,y,z)$ and coordinate time is measured by $t$.
Coordinate velocity (three-velocity) is given by $\vec{v} \equiv d_t \vec{x}$ and the individual Cartesian components of the velocity are denoted as $\vec{v} = (v_x, v_y, v_z)$.
Proper time  $\tau$ (measured with a co-moving clock) is connected to the coordinate time via the Lorentz factor $\gamma \equiv d_{\tau} t$.
The proper velocity (spatial components of the four-velocity) is $\vec{u} \equiv d_{\tau} \vec{x} = \gamma \vec{v}$.
The Lorentz factor and the velocities are connected by the expression
\be
\gamma^2 = 1 + (u/c)^2 = (1-(v/c)^2)^{-1},
\ee
where $c$ is the speed of light, $u = |\vec{u}|$, and $v = |\vec{v}|$.
Acceleration is denoted with $\vec{a} \equiv d_t \vec{u}$.

A six-dimensional phase-space density distribution for particle species s is given by $\f{s} \equiv \f{s}(\vec{x}, \vec{u}; t)$.
Thus, $\f{s}\, d^3 x \, d^3 u$ is the number of particles in the six-dimensional differential phase space volume between $\vec{x}$, $\vec{u}$ and $\vec{x} + d\vec{x}$, $\vec{u} + d\vec{u}$.

The evolution of $\f{s}$ is governed by the relativistic Boltzmann--Vlasov equation
\be\label{eq:vlasov}
\parfrac{\f{s}}{t} + \vec{v} \cdot \nabla_{\vec{x}} \f{s} + \vec{a}_{\mathrm{s}} \cdot \nabla_{\vec{u}} \f{s}  = \mathcal{C},
\ee
where $\nabla_{\vec{x}} = \frac{\partial}{\partial \vec{x}}$ and $\nabla_{\vec{u}} = \frac{\partial}{\partial \vec{u}}$ are the spatial and momentum parts of the differential operator $\nabla$, respectively.
The term on the right-hand side, defined as $\mathcal{C} \equiv \partial_t \f{s} ~|_{\mathrm{coll}}$, is the collision operator.
For the  Vlasov equation $\mathcal{C} = 0$ (i.e., the plasma is collisionless).
Acceleration of a charged particle $\vec{a}_{\mathrm{s}}$ is governed by the Lorentz force
\be
\vec{a}_{\mathrm{s}} \equiv \ud_t \vec{u} = \frac{q_{\mathrm{s}} }{  m_{\mathrm{s}} } \left(\vec{E} + \frac{\vec{v}}{c} \times \vec{B} \right)
   = \frac{q_{\mathrm{s}} }{  m_{\mathrm{s}} } \left(\vec{E} + \frac{\vec{u}}{\gamma c} \times \vec{B} \right),
\ee
where $\vec{E}$ and $\vec{B}$ are the electric and magnetic fields, $q_{\mathrm{s}}$ is the charge, and $m_{\mathrm{s}}$ is the mass of the particle of species s.

The moments of the distribution function define macroscopic (bulk) quantities of the plasma.
The zeroth moment of the distribution function $\f{s}$ defines the charge density of species s as
\be
\rho_{\mathrm{s}} \equiv  q_{\mathrm{s}} \int \f{s} \, \ud \vec{u}.
\ee
The total charge density is $\rho = \sum_{\mathrm{s}} \rho_{\mathrm{s}}$.
The first moment defines the current (charge flux) vector as
\be\label{eq:J}
\vec{J}_{\mathrm{s}} \equiv q_{\mathrm{s}} \int \f{s} \vec{v} \, \ud \vec{u} 
                      = q_{\mathrm{s}} \int \f{s} \, \frac{ \vec{u}}{\gamma} ~\ud \vec{u}.
\ee
The total current is $\vec{J} = \sum_{\mathrm{s}} \vec{J}_{\mathrm{s}}$.

\subsection{Maxwell's equations}

Evolution of electric field $\vec{E}$ and magnetic field $\vec{B}$ is governed by   Maxwell's equations.
These are  Gauss's law
\be\label{eq:divE}
\nabla \cdot \vec{E} = 4\pi \rho,
\ee
Gauss's law for magnetism
\be\label{eq:divB}
\nabla \cdot \vec{B} = 0,
\ee
Ampere's law
\be\label{eq:curlE}
\nabla \times \vec{E} = -\frac{1}{c}\parfrac{\vec{B}}{t},
\ee
and Faraday's law
\be\label{eq:curlB}
\nabla \times \vec{B} = \frac{4\pi}{c}\vec{J} +\frac{1}{c}\parfrac{\vec{E}}{t}.
\ee 
Charge conservation follows from these equations by taking a divergence of Eq.~\eqref{eq:curlB} and substituting Eq.~\eqref{eq:divE} to get
\be
\parfrac{\rho}{t} + \nabla \cdot \vec{J} = 0.
\ee

\subsection{Dimensionless equations}

We  now describe the actual dimensionless equations that are solved numerically.
A unit system similar to that used  in \tristan\ and \tristanmp \citep{buneman1993, spitkovsky2005, Sironi_2013} is used. 
The derivation and more thorough discussion of this unit system is given in  Appendix~\ref{app:units}.

Many quantities, such as  the electromagnetic fields, are defined on a lattice (mesh).
The discrete location of a point in this case is given as $\vec{x}^{(i,j,k)} \equiv \vec{x}(i,j,k) = (i\Delta x, j \Delta y, k \Delta z),$ where $i,j,k$ are the grid indices and $\Delta x$, $\Delta y$, and $\Delta z$ are the grid point distances in each Cartesian dimension.
Similarly, discretized time is given as $t^n \equiv t(n) = n \Delta t$.
Cells of the lattice are taken to have a fixed cube geometry: $\Delta x = \Delta y = \Delta z$.
The Courant (or Courant-Friedrichs-Lewy; \citealt{cfl}) number is defined as
\be
\Cfl = c \frac{ \Delta t }{\Delta x},
\ee
which represents a maximum numerical velocity of a signal traveling on the grid.
For explicit time integration schemes $\Cfl \leq 1$.

Electromagnetic fields are normalized with $B_0$ as $\Ehat = \vec{E}/B_0$ and $\Bhat = \vec{B}/B_0$.
Similarly, current density is normalized with $J_0$ as $\Jhat = \vec{J}/J_0$.
The values of these normalization factors are selected such that the numerically solved equations appear as  $\Delta x = \Delta t = 1$.
This also means that the grid indices $i$, $j$, and $k$ have the same numerical value as location $\vec{x}$. 

\subsubsection{Electromagnetic field module}\label{sect:fields}
The time evolution of electromagnetic fields is handled with a finite-difference time-domain (FDTD) method.
A Yee lattice \citep{yee1966} is used for the  $\vec{E}$ and $\vec{B}$ fields such that they are staggered both in space and in time
\begin{align}
\Ehat &= \left(
    \hat{E}_{x;\, i+\hf, j    , k    },\,
    \hat{E}_{y;\, i    , j+\hf, k    },\,
    \hat{E}_{z;\, i    , j    , k+\hf},
    \right)^{n} ,\\
\Bhat &= \left(
    \hat{B}_{x;\, i    , j+\hf,  k+\hf},\,
    \hat{B}_{y;\, i+\hf, j    ,  k+\hf},\,
    \hat{B}_{z;\, i+\hf, j+\hf,  k    },
    \right)^{n-\hf},
\end{align}
where $\Ehat$ is located in the middle of the cell sides and $\Bhat$ in the center of the cell faces. 
This makes it easy to calculate the curl of the fields in the subsequent equations.

In the time domain we update the  $\Ehat$ and $\Bhat$ fields with discrete forms of Eqs. \eqref{eq:curlE} and \eqref{eq:curlB} given as
\be\label{eq:fdE}
\Delta[ \Ehat ]_t = \Cfl \Delta[ \Bhat ]_{\vec{x}} - \hat{J}
\ee
and
\be\label{eq:fdB}
\Delta[ \Bhat ]_t =-\Cfl \Delta[ \Ehat ]_{\vec{x}},
\ee
where $\Delta[Q]_{t,\vec{x}}$ is the time differentiation or curl of a variable $Q$ without the $\Delta x$ or $\Delta t$ denominator.
The only normalization factor entering these equations is the Courant velocity $\Cfl$ since everything else is absorbed in $B_0$ and $J_0$.
There is no need to solve Eqs. \eqref{eq:divE} and \eqref{eq:divB} if the charge-conserving scheme is used together with the Yee staggering of the fields (see Appendix~\ref{app:fdtd} for details).

\subsubsection{Particle-in-cell module}\label{sect:pic}

The main purpose of the \textsc{pic} module is to update particle velocities according to the Lorentz force.
We express the spatial component of the four-velocity $u$ in units of $c$.
The particle species specifier s is omitted in this section for brevity.

The Lorentz force acting on a charged particle is
\be
m\frac{\mathrm{d} \vec{u} }{\mathrm{d} t} = q \left( \vec{E} + \frac{\vec{v}}{c} \times \vec{B} \right).
\ee
We simplify this expression again to make it  appear  unitless.
We  express the charge and mass as $q = q_0 \hat{q}$ and $m = m_0 \hat{m}$.
The right-hand side of the Lorentz force equation is then
\be\label{eq:fdU}
\Delta[ \uhat ]_t = \frac{\hat{q}}{\hat{m}} \left( \Ehat + \frac{\vhat}{\Cfl} \times \Bhat \right).
\ee
The actual velocity update is done in parts since $\Ehat$ and $\Bhat$ are staggered in time. 
As an example of particle momentum update, the relativistic Boris scheme \citep{boris1970} is presented in detail in Appendix~\ref{app:Boris}.
In addition to the particle velocity update routines, we need a scheme to interpolate the (staggered) $\Ehat$ and $\Bhat$ fields to the arbitrary particle locations.
This interpolation is typically implemented using a linear volume-averaging interpolation scheme although higher-order schemes can also be utilized. 

After the velocity update, we advance the particle's location.
The particle's position $\vec{x}$ is advanced in time with
\be\label{eq:fdX}
\Delta[\vec{x}]_t = \frac{ \vec{u} \Cfl }{ \gamma }.
\ee
The described particle propagation scheme is second order in time. 
The velocity is stored in the middle of the time steps and updated as $\vec{u}_{n+\hf} = \vec{u}_{n-\hf} + \Delta[\vec{\uhat} ]_{t, \,n}$; 
the location is stored at integer time steps and updated as $\vec{x}_{n+1} = \vec{x}_{n} + \Delta[ \vec{x} ]_{t, \,n+\hf}$.

Finally, when a charged particle moves on the grid it creates a current $\Jhat$ that induces an electric field via Eq.~\eqref{eq:curlB}.
We use the charge-conserving ZigZag current deposition scheme to deposit the current from the computational particles into the mesh \citep{Umeda_2003}.
This scheme is summarized in Appendix~\ref{app:zigzag}.

\section{Numerical implementation}\label{sect:numerics}

In this section we now discuss the general design principles and structure of the framework itself.
\runko\ is a hybrid \cpp/\python\ code:
low-level computationally intensive kernels are written in \cpp,\ whereas the high-level functionality is operated by \python\ scripts.
The design heavily relies on an object-oriented programming paradigm (supported naturally by both \cpp\ and \python), where data containers (attributes) and functions (methods) are grouped together into objects that interact with each other.
In \runko\ we call a group of these objects \textit{modules} (Sect.~\ref{sect:modules}) and the objects themselves are called \textit{solvers} (Sect.~\ref{sect:solvers}). 
The solvers are operated by different kinds of \textit{drivers} (Sect.~\ref{sect:drivers}).
This kind of hybrid use of  \cpp\ and \python\ ensures a fast and well-optimized low-level code, whereas the \python\ scripts allow for ease of use and straightforward extensibility of the code.

The low-level \cpp\ kernels rely heavily  on template metaprogramming features of modern \cpp\ (conforming currently to \cppv{14}\ standard
\footnote{%
    The \cppv{14}\ and $17$ standards are generally considered as ``transition standards'' bridging the gap between the older \cppv{11} and the newer \cppv{20}\ version.
    In the future, \runko\ will transition to fully conform to the \cppv{20} standard as soon as  sufficient HPC compiler support can be guaranteed.
    At the moment, many of the newer \cppv{20} template metaprogramming features are provided together with \runko\ itself, implemented using \cppv{11}.
}). 
In template metaprogramming a code template is provided to the compiler that then generates the actual source code based on it, given some extra information on how to fill the template parameters.%
\footnote{%
    We use the standard \cpp\ syntax to present template classes: 
    an object \texttt{A} with a template parameter \texttt{X} is given as \texttt{A<X>}.
}
Many of the implemented \cpp\ kernels take advantage of this by only having a $D$-dimensional general template of the algorithm.
This template of an algorithm is then specialized to the required spatial dimension (typically $D = 1$, $2$, or $3$) by the compiler during compile time.
This translates to a more bug-free code as typically only one implementation of the algorithm is required.

The low-level \cpp\ kernels are operated by user-friendly \python\ scripts.
Technically, the \cpp\ kernels are exposed to \python\ using the \pybind\ library \citep{pybind11}.
All the exposed \cpp\ objects behave like native \python\ objects.
No unnecessary memory copies are performed because \pybind\ manages memory pointers. 
After installing \runko\ these objects are available to \python\ by importing them from the \texttt{pyrunko} package (and \texttt{pycorgi}, as described in Sect.~\ref{sect:corgi}).
Not only can these classes be easily used from the \python\ scripts, but they also support inheritance in the case the user wants to extend and modify the objects.
This allows for fast prototyping as new classes can  first be developed in \python\ before implementing them in \cpp\ (if  better performance is needed).
These \python\ scripts also leverage \numpy\ \citep{numpy} and \scipy\ for performance and convenience, together with \matplotlib\ \citep{matplotlib} for visualization.
 
\subsection{Grid infrastructure}\label{sect:corgi}

\runko\ is built on top of the massively parallel \corgi\footnote{\url{https://github.com/natj/corgi}} \cpp\ template library (N\"attil\"a, \textit{in prep.}).
\corgi\ is a modern parallelized grid infrastructure library that provides the (spatial) simulation grid.
It is designed to be run on distributed memory systems (i.e., computing clusters and supercomputers) that require explicit inter-node communication.
All inter-node communications are handled by the \corgi\ library using the Message Passing Interface (\textsc{MPI}) library.%
\footnote{%
\url{https://github.com/natj/mpi4cpp}.
}
Hereafter, one MPI process is called a \textit{rank}, to make a distinction with a processor because a rank does not necessarily have to coincide with a processor.

\corgi\ uses a patch-based super-decomposition strategy to partition the simulation grid between different ranks.
This means that the full simulation grid is split into small (preferably continuous)  subregions called \textit{tiles}.
One tile can, for example, host a $10\times10$ subgrid in a 2D simulation.
If the complete simulation domain is composed of a $10^4 \times 10^4$ lattice, it can be tessellated with $10^3 \times 10^3$ of these tiles.

The implementation of a physical simulation code works by inheriting all the grid infrastructure methods from a \texttt{corgi::Tile<D>} template class.
This derived template class then automatically holds all the relevant physical attributes and methods needed to perform the simulation communications.
The tile object provided by \corgi\ is a template class that is parameterized with the spatial dimension $D$.
Specializing this dimension parameter to any given value sets the simulation grid dimensionality.

All the tiles are stored in a container provided by \corgi\ called a \textit{grid}.
Each program rank has its own grid object \texttt{corgi::Grid<D>} (specialized for $D$) that holds a varying number of tiles.
This allows us to decompose the grid so that each rank will only hold and process a small part of the simulation domain (i.e., set of tiles).
The grid class is responsible for providing all the spatial neighborhood information of the tiles and executing the inter-tile communications.
These include both local shared-memory intra-rank communications and global inter-rank MPI communications.
In practice, these methods are typically used to keep the halo regions of the tiles up to date.

There are no restrictions on the shape of the tile boundaries between different ranks.
This allows  more complex tile ownership to be used between ranks, which aims to maximize the volume and minimize the surface area, like a honeycomb tessellation in 2D.
This configuration then translates directly to less communication needed between ranks.

Both the \texttt{corgi::Tile} and \texttt{corgi::Grid} \cpp\ classes are exposed to \python.
Dimensionality (i.e., the $D$ template parameter) is automatically set by loading a correct subpackage from the \runko\ (and \corgi) Python libraries:
options are 
\texttt{oneD} for one-dimensional simulations ($N_x$ sized tiles; $\mathcal{N}_x$ dimensional global grid configuration),
\texttt{twoD} for two-dimensional simulations ($N_x \times N_y$ sized tiles; $\mathcal{N}_x \times \mathcal{N}_y$ dimensional global grid configuration),
and \texttt{threeD} for three-dimensional ($N_x \times N_y \times N_z$ sized tiles; $\mathcal{N}_x \times \mathcal{N}_y \times \mathcal{N}_z$ dimensional global grid configuration).
Here $N_{x,y,z}$ is the subgrid resolution of tiles and $\mathcal{N}_{x,y,z}$ is the simulation grid resolution in units of the tiles.

\subsection{Modules}\label{sect:modules}

\begin{figure}
\centering
    \includegraphics[width=0.40\textwidth]{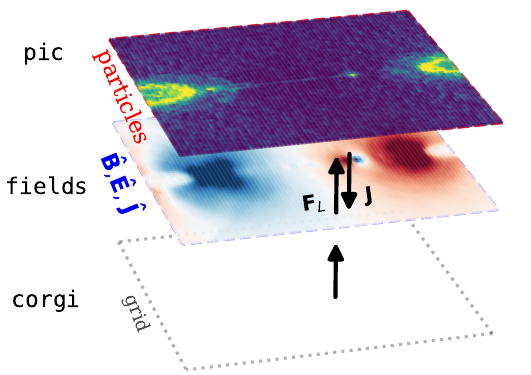}
\caption{\label{fig:code}
    Schematic presentation of the PIC code's class relationship and connectivity.
    The \textsc{fields} module is built on top of the \textsc{corgi} module that provides the spatial simulation grid.
    It contains the electromagnetic fields $\Ehat$ and $\Bhat$, and the methods to advance them in time.
    The \textsc{pic} module handles the computational particles.
    It is coupled to the \textsc{fields} module via the current $\Jhat$ induced by the moving charged particles.
    Similarly, the \textsc{fields} module is coupled to the \textsc{pic} module by the Lorentz force.
}
\end{figure}

\begin{table*}[!ht]
    \centering
    \caption{Time-advance loop of particle-in-cell module}
\begin{footnotesize}
  \begin{tabular}[c]{l l c c c}
    \hline
    \hline
      Description & Routine name & Module/Solver & Eqs. & Comments \\
    \hline
      \multicolumn{5}{c}{Grid syncronization} \\[1ex]
      Communicate $\Bhat$         & \texttt{mpi\_b0}        & \textsc{corgi} &                & global communication \\
      Communicate $\Ehat$         & \texttt{mpi\_e0}        & \textsc{corgi} &                & global communication \\
      Update halo regions         & \texttt{upd\_bc0}       & \textsc{fields}&                & local communication \\
      \hline
      \multicolumn{5}{c}{First half of $\Bhat$ advance} \\[1ex]
      Advance $\Bhat$             & \texttt{push\_half\_b1} & \textsc{fields}/\textsc{Propagator}& \eqref{eq:fdB} & $2$nd order FDTD$^\dagger$\\
      Communicate $\Bhat$         & \texttt{mpi\_b1}        & \textsc{corgi} &                & global communication \\
      Update $\Bhat$ halo regions & \texttt{upd\_bc1}       & \textsc{fields} &                & local communication \\
    \hline
  \multicolumn{5}{c}{Particle advance} \\[1ex]
      Interpolate fields to particle locations & \texttt{interp\_em} & \textsc{pic}/\textsc{Interpolator} & & Linear scheme$^\dagger$\\
      Advance particle $\vec{u}$ and $\vec{x}$ & \texttt{push}       & \textsc{pic}/\textsc{Pusher} & \eqref{eq:fdU}--\eqref{eq:fdX}; & Boris scheme:$^\dagger$ \eqref{eq:appB1}-\eqref{eq:appB2}   \\
    \hline
  \multicolumn{5}{c}{Second half of $\Bhat$ advance} \\[1ex]
      Advance $\Bhat$ & \texttt{push\_half\_b2}       & \textsc{fields}/\textsc{Propagator}& \eqref{eq:fdB} & $2$nd order FDTD$^\dagger$ \\
      Communicate $\Bhat$ & \texttt{mpi\_b2}             & \textsc{corgi} &               & global communication \\
      Update $\Bhat$ halo regions & \texttt{upd\_bc2}            & \textsc{fields} &               & local communication \\
    \hline
  \multicolumn{5}{c}{$\Ehat$ advance} \\[1ex]
      Advance $\Ehat$ & \texttt{push\_e}  & \textsc{fields}/\textsc{Propagator}  & \eqref{eq:fdE} & $2$nd order FDTD$^\dagger$\\
    \hline
  \multicolumn{5}{c}{Current $\Jhat$ calculation}    \\[1ex]
      Calculate $\Jhat$               & \texttt{comp\_curr}   &   \textsc{pic}/\textsc{Depositer} & & ZigZag scheme$^\dagger$ \\
      Initialize $\Jhat$ halo regions & \texttt{clear\_vir\_cur} & \textsc{fields} &  & \\
      Communicate $\Jhat$             & \texttt{mpi\_cur}        & \textsc{corgi} & & global communication \\
      Update $\Jhat$ halo regions     & \texttt{cur\_exchange}   & \textsc{fields} & & local communication \\
    \hline
  \multicolumn{5}{c}{Particle communications} \\[1ex]
      Tag outflowing particles    & \texttt{check\_outg\_prtcls}  & \textsc{pic} & & \\
      Pack them into a message    & \texttt{pack\_outg\_prtcls}   & \textsc{pic} & & \\
      Communicate particles       & \texttt{mpi\_prtcls}          & \textsc{corgi} & & global communication \\
      Unpack received messages    & \texttt{unpack\_vir\_prtcls}  & \textsc{pic} & & \\
      Receive incoming particles  & \texttt{get\_inc\_prtcls}     & \textsc{pic} & & local communication \\
      Remove outflowing particles & \texttt{del\_trnsfrd\_prtcls} & \textsc{pic} & & \\
      Clear global message arrays & \texttt{del\_vir\_prtcls}     & \textsc{pic} & & \\
    \hline
  \multicolumn{5}{c}{Deposit current $\Jhat$ to $\Ehat$}    \\[1ex]
      Add $\Jhat$ to $\Ehat$ & \texttt{add\_cur} & \textsc{fields} & \eqref{eq:fdE} & \\
    \hline
  \multicolumn{5}{c}{I/O and analysis}    \\[1ex]
      Write output files & \texttt{io}    & \textsc{io} & & \\
    \hline
 \end{tabular}
\end{footnotesize}
\label{tab:pic_loop}
\begin{center}
  {\small{
      Notes: The listed routines are evaluated in order from top to bottom, with one loop corresponding to one time step.
      Changeable components and algorithms are flagged with a dagger ($^\dagger$).
}}
   \end{center}
\end{table*}

The actual physics of the simulation is implemented in derived tiles that are based on the \corgi\ base tiles.
We call these tiles and all the related objects and functions  \textit{modules}.%
\footnote{%
    Technically, a module corresponds to a namespace in \cpp\ and a subpackage in \python.
}
An important part of the framework design is that different modules can be combined with polymorphic multiple inheritance (i.e., deriving methods and attributes from multiple other modules).
This allows an efficient re-use of different modules when designing a new module with similar or overlapping physics concepts.

As an example, the PIC code (Sect.~\ref{sect:pic}) uses  \texttt{corgi::Tile} as its first base class to inherit the grid infrastructure and \texttt{fields::Tile} as its second base class to inherit the electromagnetic FDTD routines (Sect.~\ref{sect:fields}).
The \textsc{pic} module itself only defines the particle containers and methods to manipulate particle movement.
The two modules are connected  by the current $\Jhat$ that is created by the moving charged particles.
This current then induces a change in the  $\Ehat$ and $\Bhat$ fields that affects the Lorentz force experienced by the particles (see Fig.~\ref{fig:code}).

\subsection{Solvers}\label{sect:solvers}

Each module can contain several  \textit{solvers} that operate on and/or change the internal attributes of the module's tiles.
Solvers are created by first defining an interface with an abstract class.
The actual solver implementations that can be instantiated (i.e., the concrete classes) can then be implemented by creating a derived class from this abstract interface.
These abstract interface classes are bound to \python\ objects via the  ``trampoline'' classes.
This automates the binding of any user-defined concrete solver implementations.
A typical requirement of any solver is that it should provide a \texttt{solve} method that takes (at least) the module's tile as an input.

As an example, the \textsc{pic} module defines  an abstract \textsc{Interpolator} solver interface.
The purpose of this solver object is to interpolate the $\Ehat$ and $\Bhat$ fields, defined on a lattice, to arbitrary particle locations.
One of the currently provided \textsc{Interpolator} instances is the \textsc{Linear} solver that implements a linear first-order interpolation scheme to approximate the field values between the grid points.
Thanks to this design pattern, it is very easy to extend the module to include other interpolation methods, such as  a second-order polynomial or a spectral interpolation.
The framework is agnostic about how the job of the solver is actually fulfilled; 
it is enough that it  conforms to the abstract interface class standard,   after which it can be automatically incorporated into the module routines.

\begin{table*}[!ht]
    \centering
    \caption{List of \runko\ modules and solvers}
\begin{footnotesize}
  \begin{tabular}[c]{l c l c c}
    \hline
    \hline
Module / Solver & Instance & Description & Eqs. & References \\[2ex]
    \hline
\textsc{fields}                       &                 & Maxwell's field equations module &             & \\[1.5ex]

\phantom{xxx}\textsc{Propagator}  &                 & Field solver              &             & \\[1ex]
                            & \texttt{FDTD2} & $2$nd order FDTD solver & \eqref{eq:fdE} and \eqref{eq:fdB} & \citet{yee1966} \\
                            & \texttt{FDTD4} & $4$th order FDTD solver &                                   & \citet{Greenwood2004} \\
                            & \texttt{FDTDGen} & FDTD solver  with free coefficients&  & \citet{Blinne_2018} \\[1ex]
\phantom{xxx}\textsc{Filter}  &                 & Current smoothing              &             & \\[1ex]
                            & \texttt{Binomial2} & $3$-point (binomial) digital filter &  & \citet{birdsall1985} \\
                            & \texttt{Compensator2} & $3$-point ``compensator'' filter &  & \textit{ibid.} \\

\hline
\textsc{pic} & & Particle-in-cell module & & \\[1.5ex]

\phantom{xxx}\textsc{Pusher}        &                 & Particle $\vec{x}$ and $\vec{u}$ propagation &                                           & \\[1ex]
                          & base class & Simple velocity Verlet propagator      & \eqref{eq:fdX}                            & \citet{Verlet_1967}\\
                              & \texttt{BorisPusher}  & Relativistic Boris pusher              & \eqref{eq:appB1}-\eqref{eq:appB2}         & \citet{boris1970}  \\
                              & \texttt{VayPusher}    & Vay pusher                             &                                       & \citet{Vay2008}    \\
                              & \texttt{HigueraCaryPusher} & Higuera-Cary pusher &  & \citet{Higuera_2017}    \\
                              & \texttt{rGCAPusher} & Guiding-center pusher &  & \citet{Bacchini_2020}    \\[1ex]

\phantom{xxx}\textsc{Interpolator}  &                 & Field interpolator to particle location    &                                           & \\[1ex]
                          & \texttt{LinearInterpolator} & Linear $1$st order interpolator            &                                           & \\
                          & \texttt{QuadraticInterpolator}& Quadratic $2$nd order interpolator       &                                           & \\
                          & \texttt{CubicInterpolator}& Cubic $3$rd order interpolator       &                                           & \\
                          & \texttt{QuarticInterpolator}& Quartic $4$th order interpolator       &                                           & \\[1ex]

\phantom{xxx}\textsc{Depositer}     &                 & Current deposition                         &                                           & \\[1ex]
                              & \texttt{ZigZag} & First order ZigZag scheme                  & \eqref{eq:appE1}-\eqref{eq:appE2}     & \citet{Umeda_2003} \\
                              & \texttt{ZigZag\_2nd} & $2$nd order ZigZag scheme    &    & \citet{Umeda_2005} \\
                              & \texttt{Esikerpov\_2nd} & $2$nd order Esikerpov scheme    &    & \citet{esirkepov2001} \\
                              & \texttt{Esikerpov\_4th} & $4$th order Esikerpov scheme    &    & \textit{ibid.} \\
    \hline
 \end{tabular}
\end{footnotesize}
\label{tab:modules}
\end{table*}

\subsection{Drivers}\label{sect:drivers}

The actual simulation codes in \runko\ are executed by the \textit{drivers}.
Drivers are short \python\ scripts that use the \cpp\ bindings of the tiles and solvers.
They typically consist of an initialization section in the beginning where tiles of some specific module are loaded into the \corgi\ grid container.
After this, the data attributes of the tiles are initialized.
The drivers also include the main time propagation loop where solvers are applied to the tiles in the correct order.
As an example, Table~\ref{tab:pic_loop} presents the main time integration loop of the PIC code.

The main advantage of using the low-level \cpp\ classes via the \python\ driver scripts is the ease of use.
For example, various complex initialization routines are rendered much easier to implement in a high-level language that also supports data visualization.
It also opens up the possibility of performing  the analysis (or at least part of it) on the fly during the simulation.

\subsection{Currently implemented modules}

The currently implemented modules include the \textsc{fields} and \textsc{pic} modules.
These and the corresponding solvers are listed in Table.~\ref{tab:modules}.

The \textsc{fields} module (Sect.~\ref{sect:fields}) is responsible for advancing the electromagnetic fields via the Maxwell's equations.
This module inherits its spatial dimensionality $D$ from the \corgi\ tiles.
Internally it has a mesh container that stores the field components as  $D$-dimensional lattices.
It also holds methods for advancing these fields in time using Eqs.~\eqref{eq:fdE} and \eqref{eq:fdB}.
The module is agnostic about the form of the incoming current; it can be coupled to any other module to provide this closure.

The \textsc{pic} module (Sect.~\ref{sect:pic}) handles the particle propagation and provides the current closure for the \textsc{fields} module (see also Fig.~\ref{fig:code}).
Attributes of the electromagnetic field lattices are inherited from the \texttt{fields::Tile}.
In addition to the spatial dimensionality template parameter $D$, the \textsc{pic} module has another template parameter for the velocity space dimensionality $V$ (i.e., \texttt{pic::Tile<D,V>)}.
By specializing the $V$ template parameter, the solvers acting on the \texttt{pic::Tile} take into account the corresponding velocity space dimensionality.
For example, setting $V=2$ equals  manipulating the velocity vector as $\vec{u} = (u_x, u_y)$, whereas $V=3$ modifies a full three-dimensional velocity vector $\vec{u} = (u_x, u_y, u_z)$.
The \textsc{pic} module also defines various solvers that operate on the \texttt{pic::Tile}s (see Table~\ref{tab:modules} for all the solvers implemented in the framework).

The final important module in the current framework is the \textsc{io} module, which is responsible for the data input and output functionality.
It is currently more detached from the physical implementations providing only solvers that act on the tiles.
These solvers can be categorized into two distinct types of input \textit{readers} and output \textit{writers}.
The input readers load data into the tiles from storage files, whereas the output writers write data from the tiles to the disk.
Currently implemented readers and writers are capable of operating on full simulation snapshot restart files using the hierarchical data format (\textsc{HDF5}).
In addition, both field and particle classes have more storage-efficient writers that reprocess the data on the fly  during the simulation and only write a smaller subsample of the full data into disk.

\section{Problem showcase}\label{sect:results}

In this section we showcase the   new PIC code implemented into the \runko\ framework.
First, we demonstrate the code validity by comparing the simulation results of the relativistic Weibel instability to the known analytical growth rates.
Next, we demonstrate the same instability on a larger scale by presenting results from an unmagnetized collisionless shock simulation. 
Finally, we measure the numerical performance of the code by using results from relativistic turbulence simulations.

\subsection{Relativistic Weibel instability}\label{sect:weibel}

\begin{figure}[t!]
\centering
    \includegraphics[trim={0.0cm 0.0cm 0.0cm 0.0cm}, clip=true, width=0.45\textwidth]{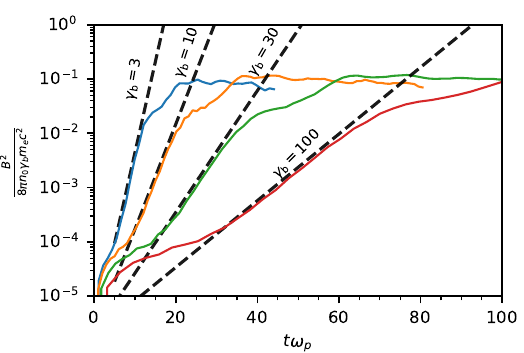}
    \caption{\label{fig:weibel}
        Simulations of  Weibel instability growth rates in relativistic counter-streaming plasmas. 
        The different curves show the magnetic field energy density $B^2/8\pi$ (normalized by the relativistic plasma enthalphy density, $\gamma_\mathrm{b} n_0 m_\mathrm{e} c^2$) as a function of time $t \omega_{\mathrm{p}}$ for beam velocities of $\gamma_{\mathrm{b}} = 3$ (blue), $10$ (orange), $30$ (green), and $100$ (red).
        The dashed lines show the expected analytic growth rates.
}
\end{figure}

We start showcasing the code capabilities by first simulating a standard astrophysics-motivated plasma instability, commonly known as the Weibel instability \citep{Weibel_1959, Fried_1959, bret2010}. 
The relativistic Weibel instability in a collisionless plasma is triggered when two counter-streaming plasma populations start to experience small charge-dependent deflections, seeded by the noise. 
The resulting modes are electromagnetic so perturbations in both $\vec{E}$ and $\vec{B}$ are important.
Therefore, the presented simulations also help us probe the validity of the solutions to the full Vlasov--Maxwell equations.

\subsubsection{Problem setup}

We simulate the linear growth phase of the instability by following two counter-streaming pair plasma populations in a 2D  simulation domain.
The plasma is modeled with $32$ particles per cell per species, and is initialized to have a small thermal spread of $\theta = kT/m_\mathrm{e} c^2 = 10^{-5}$ and to flow relativistically along $\nvec{x}$ as $\vec{u} = \pm\gamma_{\mathrm{b}} \nvec{x}$.
Electromagnetic fields are propagated in time with the standard \texttt{FDTD2} solver, current is deposited with the \texttt{ZigZag} scheme, particle interpolation is done with the \texttt{LinearInterpolator} method, and particle propagation is performed with the  \texttt{BorisPusher} solver.
The simulation domain is captured with a grid of $320 \times 80$ cells and the skin depth is resolved with $10$ cells.
We use a standard time step of $\Delta t = 0.45 \Delta x$.
We use symmetric beams with Lorentz factors of $\gamma_{\mathrm{b}} = 3$, $10$, $30$, and $100$.

\subsubsection{Comparison to analytical estimates}

The analytic linear growth rates of the Weibel instability are known for symmetric beams.
\cite{bret2010} presents the filamentation instability growth rate $\delta$ as
\begin{equation}
    \frac{\delta }{\omega_{\mathrm{p}}} = \beta_\mathrm{b} \sqrt{ \frac{2}{\gamma_\mathrm{b} }},
\end{equation}
where $\gamma_\mathrm{b} = (1-\beta_{\mathrm{b}}^2)^{-1/2}$ is the bulk Lorentz factor of the beams, $\omega_\mathrm{p} = \sqrt{4\pi n_\mathrm{e} e^2/\gamma_\mathrm{b} m_\mathrm{e}}$ is the plasma frequency, and  $n_\mathrm{e}$ is the plasma number density (of electrons).
We  compare the simulation results to these theoretical estimates in Fig.~\ref{fig:weibel} and find the linear instability growth rate to be perfectly matched with the expectations.

These simulations are not particularly demanding and can therefore serve as a helpful introduction to using the code.
The discussed setup is also general enough that it could be easily reformulated to study other analytically intractable plasma beam instabilities.

\subsection{Formation of collisionless shocks}\label{sect:shocks}

\begin{figure*}[t!]
\centering
    \includegraphics[trim={1.0cm 2.0cm 0.0cm 3.0cm}, clip=true, width=0.85\textwidth]{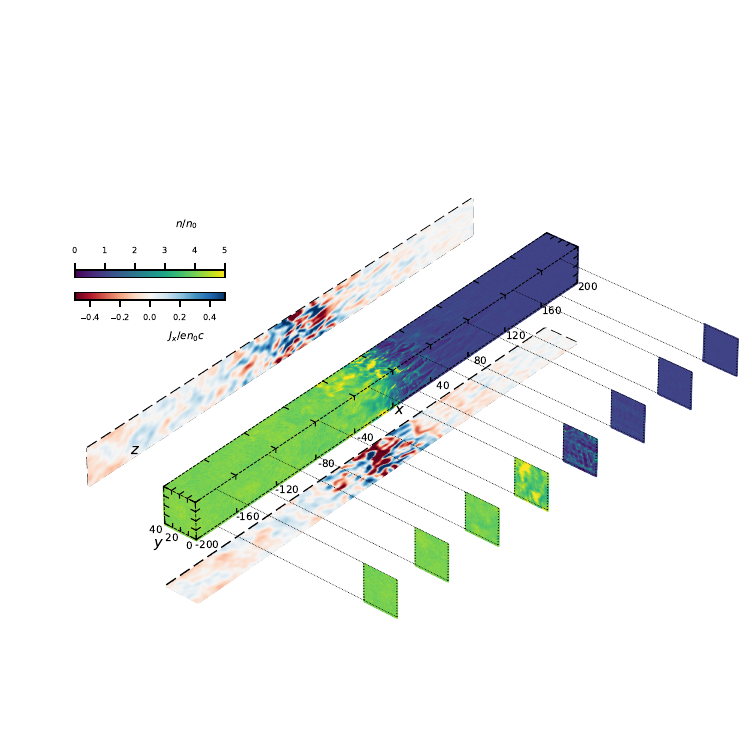}
\caption{\label{fig:shock}
    Visualization of a relativistic collisionless shock in unmagnetized pair plasma at $t\omega_\mathrm{p} \approx 500$.
    The box periphery shows the plasma number density $n/n_0$, and the exploded panels the electric current density along the flow $J_x/e n_0 c$.
    The shock is located at the middle of the box at $x = 0$;
    the same region shows strong current filaments that are responsible for mediating the shock.
}
\end{figure*}

The relativistic Weibel filamentation setup can be expanded to a realistic 3D simulation and used to model the plasma physics of collisionless shocks.
The shock is formed by the growth and nonlinear saturation of the same instability.

We find  this setup to be one of the most demanding numerical tests for PIC codes because the numerical algorithms need to be able to both sustain the relativistic upstream bulk flow and correctly model the kinetic shock physics. 
The upstream is unstable to the  numerical Cherenkov instability \citep{Godfrey_1974} and is, in addition, sensitive to any numerical error in the current deposition;
even tiny perturbations in the relativistically advected upstream plasma can trigger an isotropization of the bulk flow.
The shock front, on the other hand, is sustained by a strong electric current that is sensitive to charge conservation.
We use this setup to stress test the full 3D PIC routines and their correctness.

\subsubsection{Problem setup}

The numerical setup consists of a elongated shock ``tube'' along the $\nvec{x}$-axis and a reflecting wall on the left side of the domain.
Initially, we set all the plasma in the domain to move from right to left with a velocity $\vec{u} = -\Gamma \beta \nvec{x}$.
The plasma is reflected from the left simulation domain and forms a counter-propagating beam.
The beam is unstable to the Weibel instability that will quickly stop the right-moving component.
This causes the  formation of a collisionless shock. 

We simulate the shock in an elongated box with a size $L_\parallel \times L_\perp^2 = 512 \times 64^2 (c/\omega_\mathrm{p})^3$ using ten cells per skin depth so that the full domain is covered with a grid of $5120 \times 640^2$.
The plasma  flows with a relativistic velocity $\Gamma = 10$ and is modeled with 64 particles per cell per species.
Electromagnetic fields are propagated in time with the fourth-order \texttt{FDTD4} scheme.
The resulting current is calculated using the \texttt{ZigZag} scheme, particle interpolation is done via the \texttt{LinearInterpolator}, and the current is smoothed with eight passes of the binomial filter (\texttt{Binomial2}).
We use a time step of $\Delta t = 0.45 \Delta x$.
The pair plasma has an initial thermal spread of $\theta = kT/m_\mathrm{e} c^2 = 2\times10^{-5}$ and is updated using the \texttt{HigueraCaryPusher}.
We start the simulation with no background magnetic field, which corresponds to an unmagnetized shock.

The plasma is injected into the domain from a receding right wall that moves at the speed of light, from left to  right;
this enables longer simulations both because the particles are not ``alive'' as long as they normally would be (suppressing numerical Cherenkov instability) and by saving computational time (since we do not need to initially propagate all the particles in the domain).
The shock front is adaptively tracked on the fly, and only the high-resolution grid data around it is saved.
The grid is decomposed into cyclic MPI domains of length $1280$ cells in the $\nvec{x}$ direction (i.e. five cycles per the full box);
this results in a more even memory load for the simulation.

\subsubsection{Physical interpretation and numerical tests}

A collisionless shock is quickly formed when the simulation is started.
The shock is mediated by a formation of Weibel filaments that grow, saturate, and merge; 
this stops the counter-streaming plasma (see Fig.~\ref{fig:shock}).

The simulated shock structure and dynamics are  well described by the theoretical expectations and previous numerical experiments \citep[e.g.,][]{sironi2009, Plotnikov_2018}.
The filaments form and induce the shock in about $100 \omega_\mathrm{p}^{-1}$.
The upstream flow is compressed into a  denser downstream with a compression ratio $n_{\mathrm{down}}/n_{\mathrm{up}} \approx 4.1 \pm 0.15$, and the shock propagates into the upstream with a mildly relativistic velocity, $\beta_{\mathrm{sh}} \approx 0.32 \pm 0.03$ (measured for $200 \le t\omega_\mathrm{p} \le 1000$).
Unmagnetized MHD jump conditions, in comparison, predict, $n_{\mathrm{down}}/n_{\mathrm{up}} = 1 + (\Gamma + 1)/[\Gamma(\gamma_{\mathrm{AD}}-1]) = 4.3$ for the compression ratio, and $\beta_{\mathrm{sh}} = (\gamma_{\mathrm{AD}}-1)(\Gamma-1)/\Gamma) = 0.3$, given the 3D adiabatic index $\gamma_{\mathrm{AD}} = 4/3$ \citep[see, e.g.,][]{Plotnikov_2018}.
The shock ramp itself has a length of roughly $\approx 50 c/\omega_\mathrm{p}$ and is accommodated by an increase in the electric field energy density just ahead of the moving front.

We  also specifically tested that the upstream remains smooth and stable for an extended period of time $\gtrsim 1000 \omega_\mathrm{p}^{-1}$ with electric and magnetic field fluctuations of $\delta E$, $\delta B \lesssim 10^{-6}$ (in absolute code units; the value corresponds to  floating-point accuracy).
This is a stringent test that can be used to verify that the particle-to-grid and grid-to-particle routines are all correctly implemented, and no charge or electric current is lost;
in the opposite case, any numerical error will quickly grow and isotropize the flow. 
Similarly, the formation of the shock front supports the notion that all the 3D routines are correct.
We note that the shock simulation is a particularly demanding code test because of the above details.

\subsection{Decaying relativistic kinetic turbulence}\label{sect:turb}

\begin{figure*}
\centering
    \includegraphics[width=0.48\textwidth]{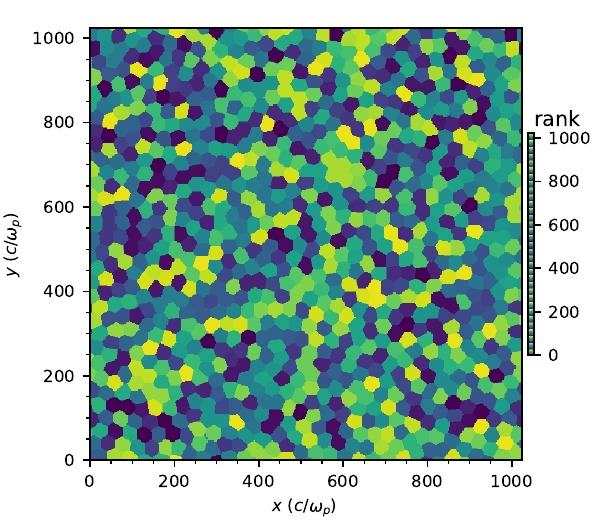}
    \includegraphics[width=0.48\textwidth]{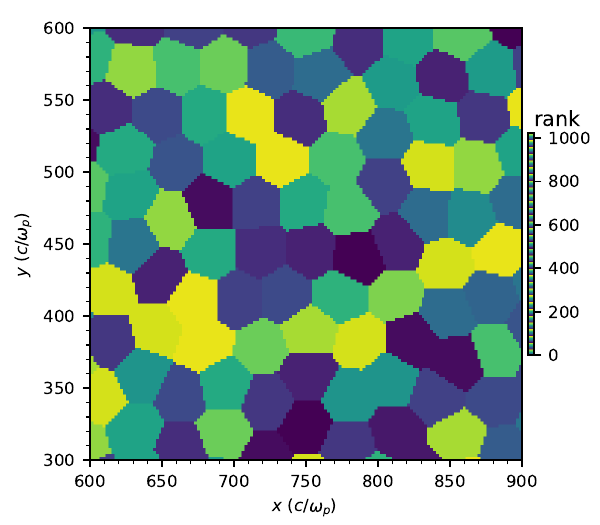}
    \includegraphics[width=0.48\textwidth]{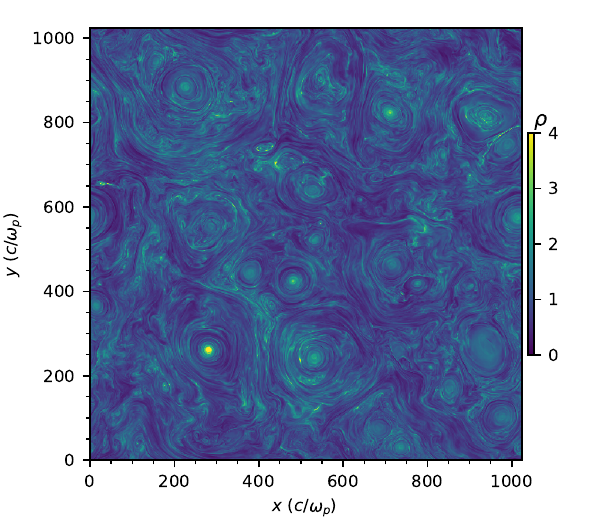}
    \includegraphics[width=0.48\textwidth]{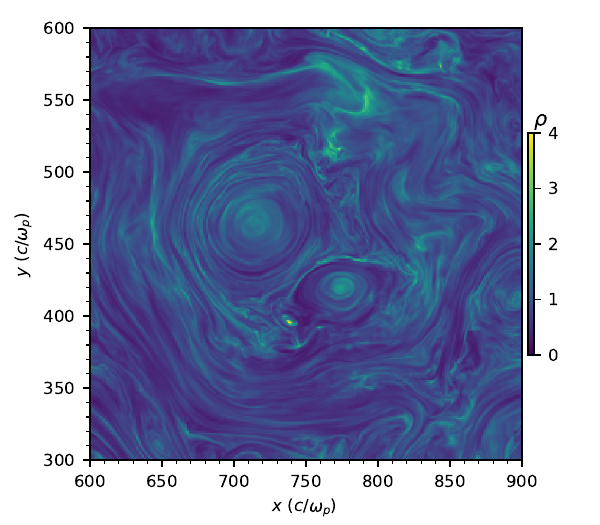}
\caption{\label{fig:turb}
    Visualizations of the kinetic turbulence simulations.
    The top row shows the honeycomb-like domain decomposition that partitions the grid between different processors with the \corgi\ library (the color map is cyclic).
    The bottom row shows the plasma density (in units of the initial plasma mass density $\rho_0$).
    The left panels show the full simulation domain, whereas the right panels focus on a smaller subregion that better illustrates the small-scale structures.
}
\end{figure*}

In this section we use a  relativistic decaying turbulence setup to showcase the code performance and scalability metrics.
The setup is presented and discussed in detail in \citet{Comisso_2018, Comisso_2019, Nattila_2021}.
Similar driven turbulence simulations are performed in \citealt{Zhdankin_2017b, Zhdankin_2017a}, among others.

\subsubsection{Problem setup}

We consider a uniform pair plasma that is suddenly stirred.
Due to this large-scale driving motion, a turbulent cascade quickly develops.
We study the  2D3V setups, where the cascade forms in the planar $x-y$ direction and is captured by the spatial 2D simulation domain.
All three velocity components are considered, however,  including the out-of-the-plane $z$ direction.
To physically mimic confinement of the cascade to a 2D plane, a guide field is imposed in the $z$ direction as $\vec{B}_0 = B_0 \hat{\vec{z}}$ ($\hat{\vec{z}}$ is the unit vector along the  $z$-axis).

Presence of the strong guide-field renders the plasma magnetically dominated.
The plasma magnetization parameter we use is
\be
\sigma = \frac{B_0^2}{4\pi n_0 \bar{\gamma} m_\mathrm{e} c^2},
\ee
where $n_0$ is the initial number density and $\bar{\gamma}$ is the mean thermal Lorentz factor of the plasma.
Here we focus only on a strongly magnetized case of $\sigma = 10$.
Initially, the plasma is defined to be warm with a temperature of $\theta = kT/m_\mathrm{e} c^2 = 0.3$ ($\bar{\gamma} \approx 1.6$).

The spatial resolution of our fiducial simulation is $5120^2$ cells ($512 \times 512$ tiles with internal tile resolution of $10\times10$).
The initial plasma  skin depth $c/\omega_\mathrm{p}$ is resolved with five cells.
This determines our box size as $L \approx 1000 c/\omega_\mathrm{p}$.
We use $64$ particles per cell per species,  totaling  $128$ particles per cell for both electron and positron species.
In total, we then propagate $\sim\!10^{9}$ particles in the simulation.

The sudden large-scale stirring is mimicked by perturbing the planar magnetic field $\vec{B}_{\perp}$ with strong $\langle B_{\perp}^2 \rangle \equiv \delta B^2 \sim B_0^2$ uncorrelated fluctuations in the  $x$ and $y$ directions following \citet{Comisso_2018}.
This drives strong turbulence with semi-relativistic bulk velocities. 
The forcing modes, $m,n \in \{1,\ldots,8\}$, are initialized with equal power as
\begin{align}
    \delta B_x &= \phantom{+} \sum_{m,n} \beta_{mn} n \sin(k_mx+\phi_{mn})\cos(k_ny+\psi_{mn})  ,\\
    \delta B_y &= -\sum_{m,n} \beta_{mn} m \cos(k_mx+\phi_{mn})\sin(k_ny+\psi_{mn}),
\end{align}
where $k_m = 2\pi m/L$ and $k_n = 2\pi n/L$ are the wave numbers along the  $x$ and $y$ dimensions, respectively. 
The forcing is uncorrelated so $\phi_{mn}$, $\psi_{mn} \in [0,2\pi[$, correspond to random phases.
The forcing amplitude is 
\begin{equation}
\beta_{mn} = 2\frac{\delta B}{N\sqrt{m^2 + n^2}},
\end{equation}
where $N=8$ in our case.
The largest forcing mode ($k_N=2\pi N/L$) defines the energy-carrying scale and the characteristic size of the largest magnetic eddies $l_0 = 2\pi/k_N$.

\subsubsection{Numerical efficiency and parallel scaling}

\begin{figure}[t!]
\centering
    \includegraphics[width=0.48\textwidth]{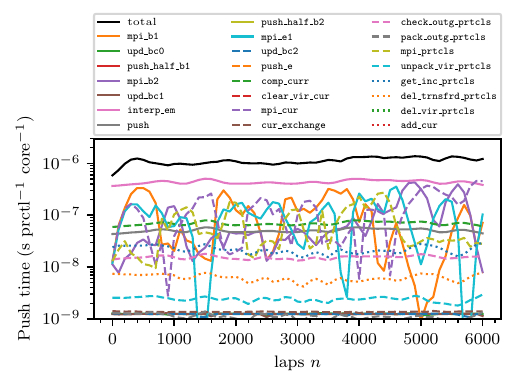}
\caption{\label{fig:perf_analysis}
Code performance analysis for $5120^2$ resolution turbulence runs with $1024$ cores.
The black line shows the total push time;  the other curves show the individual results for each component in the loop (see Table~\ref{tab:pic_loop}).
    In a perfectly load-balanced state ($n \lesssim 200$) the total particle push time is measured to be around $0.7~\mu\mathrm{s}$, whereas in a strongly load-imbalanced state an average time of about $1.1~\mu\mathrm{s}$ is obtained.
}
\end{figure}

The turbulence cascade is formed as the plasma relaxes from the non-equilibrium state (Fig.~\ref{fig:turb}, bottom row; see, e.g., \citealt{Nattila_2021} for a more in-depth physical interpretation).
We use the turbulence simulation setup to probe the numerical performance and parallel scaling of the code.

This setup is advantageous not only because it is a real physical problem, but also because we can probe both the  load-balanced and load-imbalanced performance of the PIC code. 
Initially the simulations are in the load-balanced state (computing time being dominated by particle push time), but after the cascade forms we observe a strong numerical imbalance between different ranks because of the strong fluctuations in particle number densities (computing time dominated by communication routines).
No dynamic load-balancing algorithm is used.
Here we mainly focus on these transition scales of around $\sim10^3$ processors where both of these regimes can be simulated, leaving extremely large simulations with $\gtrsim 10^5$ processors and dynamic load balancing for future work.
The scaling measurement reported here is performed in the Kebnekaise computing cluster.%
\footnote{
    Kebnekaise is a Lenovo cluster with $19288$ cores (mainly Intel Xeon E5-2690v4 Broadwell CPUs with a $2.6~\mathrm{GHz}$ base frequency) and $64~\mathrm{GB}$ of memory per each NUMA island in a node, and has FDR Infiniband switches for communication.
}
We use the second-order FDTD solver (\texttt{FDTD2}), Boris pusher (\texttt{BorisPusher}), linear interpolation (\texttt{LinearInterpolator}), and first-order ZigZag scheme (\texttt{ZigZag}) to perform the simulations.
Calculations are done using the double floating-point precision.

Good numerical performance is obtained for the PIC code.
In an ideal completely load-balanced state the average total particle update time per processor is  around $0.7~\mu\mathrm{s}$ (see Fig.~\ref{fig:perf_analysis} around $n \lesssim 200$).
In the strongly imbalanced simulation state (dominated mainly by inter-rank particle communications) the particle update time per processor is  around $1.1~\mu\mathrm{s}$ when the turbulent cascade has developed (Fig.~\ref{fig:perf_analysis} for $n \gtrsim 4000$).
The decrease in speed is caused by an increase in the evaluation time of routines communicating the fields and particles to the neighboring ranks.

We also monitor all the separate routines (see Table~\ref{tab:pic_loop}) independently to probe the numerical performance of the code in more detail.
We note, however, that this monitoring is only done for one particular rank so the component-wise evaluation times reported here might fluctuate  between different ranks experiencing differing loads.
However, the numbers are indicative of where most of the evaluation time is spent.
As seen in Fig.~\ref{fig:perf_analysis}, in the ideal case most of the evaluation time (around $0.3~\mu\mathrm{s}$ per particle per processor) is spent in the interpolation routine (\texttt{interp\_em}; pink solid line) where $\Bhat$ and $\Ehat$ are computed at the particle locations between the grid points.
The large cost of the  interpolation step is caused by frequent cache misses. The
interpolation routine consists of many random access operations of the field arrays because particles are not sorted by their location;
this prevents the processor cache prefetching from correctly predicting the incoming requested values.
The second most expensive routine is the current calculation (\texttt{comp\_curr}; green dashed line) with a typical evaluation time of $0.06~\mu\mathrm{s}$ per particle per processor.
It has a similar problem with the cache memory because writing the current to the array is preformed with an unpredictable access pattern.
When communication costs increase, the MPI messaging routines start to be as costly as the current deposition with an average time of about $0.1~\mu\mathrm{s}$ per particle per processor.

\begin{figure}[t!]
\centering
    \includegraphics[width=0.48\textwidth]{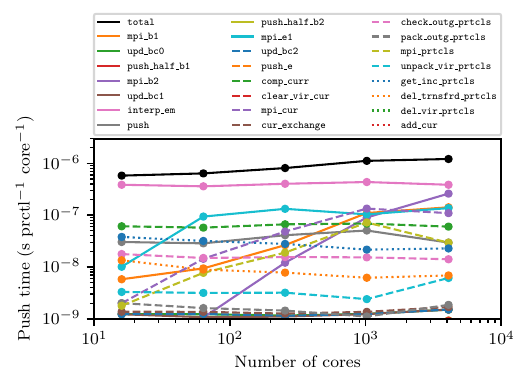}
\caption{\label{fig:wscaling}
    Weak scaling results of the PIC code measured in a strongly load-imbalanced simulation state.
    Scaling is presented in terms of the mean particle push time per processor for various simulations that have a fixed ratio of total number of particles to processors.
    The black line shows the total mean push time;   the other curves show the  individual results for each component in the loop (see Table~\ref{tab:pic_loop}).
    Ideal scaling corresponds to a horizontal line, although there is a slight increase in the evaluation time due to more time spent on communication routines as the number of processors increases.
}
\end{figure}

\begin{figure}[t!]
\centering
    \includegraphics[trim={0cm 0.0cm 0cm 0.4cm}, clip=true, width=0.48\textwidth]{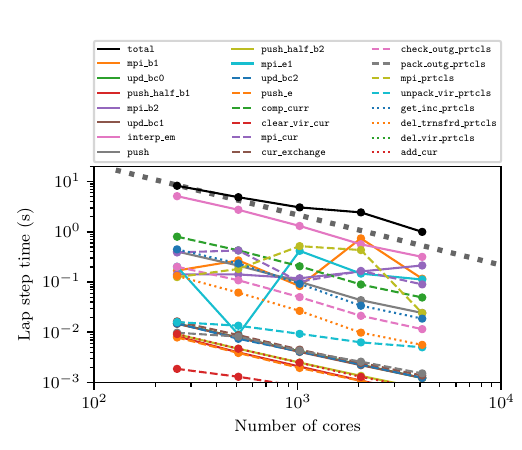}
\caption{\label{fig:sscaling}
    Strong scaling results of the PIC code measured in a strongly load-imbalanced simulation state.
    Results are for a $5120^2$ box size with $128$ particles per cell.
    The black line shows the total evaluation time per lap; the  other curves show the  individual results for each component in the loop (see Table~\ref{tab:pic_loop}).
    The ideal scaling corresponds to the thick black dotted line.
}
\end{figure}

The code shows good weak (Fig.~\ref{fig:wscaling}) and strong scaling (Fig.~\ref{fig:sscaling}) when the simulations transform from cpu-bound to communication-bound.
Weak scaling proceeds as $1.0 : 1.10 : 1.40 : 1.93 : 2.09$ for every quadrupling of the  number of processors $16: 64 : 256 :1024 : 4096$.
Strong scaling proceeds as $1.0 : 1.18 : 1.48 : 2.36 : 1.93$ against an ideal scaling for every doubling of the number of processors  $256 : 512 : 1024 : 2048 : 4096$.
The increase in the evaluation time when the number of processors is increased can indeed be attributed to the increase in the evaluation time of the communication routines (field updates: \texttt{mpi\_b1}, \texttt{mpi\_b2}, \texttt{mpi\_e1}, \texttt{mpi\_cur}, \texttt{clear\_vir\_cur}; and particle updates: \texttt{mpi\_prtcls}, \texttt{unpack\_vir\_prtcls}, \texttt{del\_vir\_prtcls}).
A similar test performed in the load-balanced state yields almost ideal scaling up to $\sim10^3$ processors (scaling of roughly $1.10$ of the ideal) with a total particle update time of about $0.7~\mu\textrm{s}$.

Finally, we note that the extra costs originating from the intra-rank updates of the tile boundaries are negligible.
This extra cost is introduced because of the patch-based grid infrastructure where the smallest unit of calculation is one tile ($10\times10$ lattice with about $10^4$ particles per species per tile in this particular case).
Therefore, some communication is always needed to keep these individual units of calculation in sync (field update routines: \texttt{upd\_bc0}, \texttt{upd\_bc1}, \texttt{upd\_bc2}, \texttt{cur\_exchange}; and particle update routines: \texttt{check\_outg\_prtcls}, \texttt{pack\_outg\_prtcls}, \texttt{get\_inc\_prtcls}, \texttt{del\_trnsfrd\_prtcls}), even in systems completely relying on shared-memory parallelism.
These routines are, however, typically $10$--$100$ times cheaper than the costs of the field interpolation or MPI communication tasks.

\section{Discussion}\label{sect:discussion}

\subsection{Computational advantages}

\runko\ is a modern numerical toolbox tailored for astrophysical plasma simulations.
The framework is designed using recent computer science methods that rely on multiple levels of code abstraction.
These in turn, help to create a general and easily extensible code framework.
Furthermore, in order to encourage all kinds of  use, \runko\ is provided as an open-source software for the community.

The code is written as a hybrid \cpp/\python\ program.
The low-level \cpp\ kernels are currently designed to conform to a \cppv{14} standard with the benefit of a plethora of modern computational methods. 
For example, we heavily rely on recent \cpp\ template metaprogramming features to simplify the design of the code.
This ensures a truly modern numerical code that will not be outdated immediately upon publication.

In addition to the underlying numerically efficient \cpp\ implementation, all the classes have a \python\ interface.
This allows the user to take full advantage of all the benefits of using a high-level language without sacrificing the actual run-time speed of the code.
In this way, many of simulation setups can be quickly prototyped in a laptop and then scaled up to supercomputer platforms for the actual production runs.
Furthermore, it allows designing and implementing very complex initial conditions for simulations because the initial routines can be created using \python.

The modular design allows \runko\ to function as a fully agnostic framework for various different modeling formulations of simulating plasma with computers.
The physics is implemented via the  modules, which in practice often solve some partial differential equations or propagate Lagrangian particle species.
This means that we are not locked into  one predefined theoretical formalism of describing the evolution of the plasma.
The high degree of modularity also allows the user to easily change the numerical solvers and algorithms depending on the problem at hand.

The implementation relies on modern \cpp\ features such as polymorpishm and template metaprogramming to simplify the code and make it easily extensible.
Different solvers inherit their standard interface from an abstract base class implementation so that introduction and implementation of new physical algorithms is straightforward.
User-definable dimensionality of different solvers is handled by template specializations.
Additionally, most of the solvers are agnostic to the actual floating-point accuracy (i.e., single- or double-length floating-point precision) because the working precision is typically also specified as a template variable.

Another technical advantage of the framework is the patch-based domain decomposition strategy that is built into the grid infrastructure routines.
As shown by the performance analysis, the cost of the additional updates needed to keep the individual tiles in sync are negligible in comparison to the evaluation costs of the physics in  modules.
Having a larger number of moderately sized tiles is found to be faster than partitioning the domain with fewer but larger tiles.%
\footnote{For one particular test we  partitioned a $3D$ PIC shock simulation grid of $7680\times120^2$ with tiles of sizes $8^3$, $15^3$, or $30^3$ (keeping the number of processors fixed at 1280 cores) and found that the absolute wall-clock time per lap varies as $1.34$, $0.66$, and $1.22 \mathrm{s}$. 
Smaller but more numerous tiles are found to have more expensive communication routines, whereas larger tiles have worse solver performance. 
The optimal  size  in this particular case is for tiles  somewhere around  $15^3$.
}
The benefits of this system are that  it automatically introduces some level of load-balancing into the simulations (each rank has $\sim10^2$ tiles so load imbalances at the level of one tile are effectively smoothed out), it helps in data cache locality (array sizes and particle containers remain small and easily movable in memory), and it simplifies the global communication routines (since these communication routines are anyway needed). 
Another benefit is the possibility to optimize the MPI domain boundaries with more complex edge shapes.
An interesting example of this is the honeycomb-like tile memory configuration (see Fig.~\ref{fig:turb}, top row).
We also note that a similar decomposition technique currently seems to be the  state-of-the-art technical choice of many other recent high-performance PIC codes, such as \textsc{VPIC} \citep{Bowers_2008} and \textsc{Smilei} \citep{Derouillat_2018}.

\subsection{Future directions}

The presented framework offers a unique possibility to start exploring and experimenting with multiphysics simulations.
These are new kinds of numerical simulations that evolve multiple physical formalisms simultaneously, or in an adaptive fashion selecting the most accurate (or relevant) solver on the fly.

These heterogeneous simulations will enable us to simulate larger, and therefore more realistic systems.
In the future this can enable, for example, novel plasma simulations where the majority of the simulation domain is modeled with a force-free electrodynamic description of the plasma \citep{Komissarov2002}.
However, in some small regions with extreme conditions (e.g., current sheets, shocks) the simulation can be adaptively refined (on the fly) to the fully kinetic description that enables a realistic in situ modeling of particle acceleration.
Another possibility could be to use a multiple-domain--multiple-physics approach, where, for example, some fixed part of the simulation is described with a kinetic description, whereas some other part is modeled with MHD.
Here the domains could be divided based on some strong density gradient in the problem setup, like those found in systems with diluted nonthermal particles on top of a denser fluid-like plasma; 
physically these can model astrophysical systems like accretion disk coronas.

Another important focus is the code performance and scalability.
In practice, this mostly means minimizing the global communication costs.
One possibility of decreasing the communication costs is the use of more complex hybrid parallelization schemes.
As an example, using the patch-based domain super-decomposition encourages the use of a task-based hybrid parallelization scheme, for example similar  to the \textsc{Dispatch} framework \citep{Nordlund_2018}.
Since most of the updates of the tiles are numerically independent, these operations can be easily performed in parallel with shared-memory parallelization strategies.
This, in turn, allows  the number of tiles to be increased per MPI rank, which acts as an extra buffer against sudden load-imbalance fluctuations.
This  strategy also allows  interleaved global (non-blocking) communications and normal solver updates to be performed simultaneously.
This allows us to hide almost all the (MPI) communications since we can prioritize the evaluation of tiles such that the boundary tiles are always computed first and then sent to the neighbors, whereas the calculation of the inner tiles are continued independently by other threads while waiting for the halo tiles.
We are also currently experimenting with dynamical load balancing methods where the ownership of the tiles  changes during the simulation depending on the computational load.

Finally, since \python\ is becoming the new lingua franca of astronomy and astrophysics, it is important to note that problem initialization in \runko\ can be done purely via \python\ scripts.
This means that new users are immediately familiar with the basic approach of the code,  and  that initialization of very complex simulation setups can be done by a high-level language.
We hope that this will accelerate the use of plasma simulations to study many yet unresolved astrophysical problems.

\section{Summary}\label{sect:summary}

We started by reviewing the kinetic plasma theory and the numerically solved Vlasov--Maxwell system of equations (Sect.~\ref{sect:theory}).
The numerical implementation of the framework is discussed in Sect.~\ref{sect:numerics}.
We focused in particular on the design of the code, and introduced the different numerical components of the framework.
First, we discussed the \corgi\ grid infrastructure and the related patch-based domain super-decomposition strategy where the computational grid is partitioned into separate \textit{tiles} (Sect.~\ref{sect:corgi}).
Second, we presented the physical building blocks of the code, the  \textit{modules} (Sect.~\ref{sect:modules}).
Different physical formulations of modeling the evolution of the plasma and  electromagnetic fields are encapsulated in different modules.
Third, each module can contain several numerical algorithms called \textit{solvers} that are short \cpp\ classes that operate and modify the content of different tiles (Sect.~\ref{sect:solvers}).
Finally, these \cpp\ solvers are applied to the tiles and operated by \python\ scripts called \textit{drivers}, ensuring easy of use (Sect.~\ref{sect:drivers}).

As our first task, we  implemented a new particle-in-cell module into the framework.
The implementation is discussed in detail in Sects.~\ref{sect:fields} and \ref{sect:pic}.
The PIC code demonstrates good numerical performance with a mean particle push time per processor of about $0.7~\mu\mathrm{s}$ in an ideal load-balanced state and $1.1~\mu\mathrm{s}$ in a strongly load-imbalanced state.
Furthermore, the code is shown to scale well up to $\sim 10\,000$ processors.

We showcased the use of this PIC code by three different complex astrophysically motivated plasma simulations.
First, we simulated the linear growth and nonlinear saturation of the Weibel instability in a relativistically streaming pair plasma.
Second, we performed a mock simulation of a current-filamentation-mediated relativistic collisionless shock in full 3D.
Last, we tested the code performance by modeling a decaying relativistic kinetic turbulence in magnetized pair plasma. 
These simulations demonstrated the ease of use, extendability, and flexibility of the code.
All the setups are commonly used to study astrophysical phenomena, and can therefore serve as a useful starting point for new users.

\section*{Acknowledgments}
JN would like to thank Andrei Beloborodov, Axel Brandenburg, Luca Comisso, Troels Haugb{\o}lle, Dhruba Mitra, Åke Nordlund, Lorenzo Sironi, Anatoly Spitkovsky, and Alexandra Veledina for stimulating discussions on numerics and plasma simulations. 
We would also like to thank the referee for many helpful comments and suggestions.
The simulations were performed on resources provided by the Swedish National Infrastructure for Computing (SNIC) at PDC and HPC2N.

\bibliographystyle{aa}

\clearpage
\onecolumn
\begin{appendix}

\section{Non-dimensionalization of the Vlasov--Maxwell equations}\label{app:units}

Following \citet{jackson1975}, the Maxwell equations in an arbitrary system of units can be written as
\begin{align}
    \nabla \cdot \vec{E} &= 4\pi k_1 \rho ,\\
    \parfrac{\vec{E}}{t} &= \frac{c^2}{k_2} \nabla \times \vec{B} - 4\pi k_1 \vec{J} ,\\
    \nabla \cdot \vec{B} &= 0 ,\\
    \parfrac{\vec{B}}{t} &= - k_2 \nabla \times \vec{E},
\end{align}
where $k_1$ and $k_2$ are constants that define the unit system. 
For a Gaussian system we have $k_1 = 1$ and $k_2 = c$, whereas for rationalized MKSA $k_1 = 1/4\pi\epsilon_0$ and $k_2 = 1$.
Here $\epsilon_0$ is the vacuum permittivity (and $\mu_0$ is the vacuum permeability found from the relation $c^2 = 1/(\epsilon_0 \mu_0)$).

In this appendix we  present the unit system in use \citep[originally by][and described in detail also in the  \textsc{Tristan-MP} user-manual by A. Spitkovsky]{buneman1993}.
We select the Gaussian system here by setting $k_1 = 1$ and $k_2 = c$.
Our Maxwell equations are then simplified to
\begin{align}
    \nabla \cdot \vec{E} &= 4\pi\rho ,\\
    \parfrac{\vec{E}}{t} &= \phantom{+} c \nabla \times \vec{B} - 4\pi\vec{J} ,\\
    \nabla \cdot \vec{B} &= 0 ,\\
    \parfrac{\vec{B}}{t} &= -           c \nabla \times \vec{E},
\end{align}
and the fields appear symmetrical as  $\vec{E} \leftrightarrow \vec{B}$.
The Lorentz force in this unit system is
\be
\derfrac{\vec{u}}{t} = \frac{q}{m} \left(\vec{E} + \frac{\vec{v}}{c} \times \vec{B} \right).
\ee

We  next normalize all the variables with some fiducial values.
The most peculiar of these is our selection of distance and time scalings:
we express the distance in units of the grid spacing, $\vec{x} = \xhat \Delta x$, and time in units of the time step, $t = \that \Delta t$.
The coordinate-velocity is then $\vec{v} = \vhat \Delta x/\Delta t = \vhat c/\Cfl$ and the four-velocity is $\vec{u} = \uhat c/\Cfl = \vhat \gamma(\vec{v}) c/\Cfl$.
The fields are scaled with a fiducial field $B_0$ such that $\vec{E} = \vec{\hat{E}} B_0$ and $\vec{B} = \vec{\hat{B}} B_0$.
Similarly, the charge, mass, and current density are also presented such that $q = \hat{q} q_0$, $m = \hat{m} m_0$, and $\vec{J} = J_0 \Jhat$, respectively.
This way, all of our numerical quantities are denoted with a hat.

In numerical (code) units the equations that are being solved are then
\begin{align}
    \Delta[ \vec{\hat{E}} ]_t &=
    \phantom{+} c \frac{\Delta t}{\Delta x} \Delta[ \vec{\hat{B}} ]_{\vec{x}} - \frac{4\pi J_0}{B_0} \Delta t  \Jhat
    = \Cfl \Delta[ \vec{\hat{B}} ]_{\vec{x}} -  \vec{\hat{J}} \label{appB1} ,\\
    \Delta[ \vec{\hat{B}} ]_t &=
    - c \frac{\Delta t}{\Delta x} \Delta[ \vec{\hat{E}} ]_{\vec{x}} 
    = -\Cfl \Delta[ \vec{\hat{E}} ]_{\vec{x}} ,\\
    \Delta[ \uhat ]_t &= 
    \frac{\hat{q}q_0 \Cfl}{\hat{m}m_0 c} B_0 \Delta t \left(\vec{\hat{E}} + \frac{\vec{\hat{v}} c}{c \Cfl} \times \vec{\hat{B}} \right) 
    = \frac{\hat{q}}{\hat{m}} \left(\vec{\hat{E}} + \frac{\vec{\hat{v}}}{\Cfl} \times \vec{\hat{B}} \right), \label{appB3}
\end{align}
where in the last steps we  define
\begin{align}
    B_0 &=  \frac{m_0 c}{q_0 \Cfl \Delta t} = \frac{m_0 c}{q_0 \Cfl} \left( \frac{c}{\Cfl \Delta x} \right), \label{appB13}\\
    \frac{J_0}{B_0} &= \frac{1}{4\pi \Delta t}, \label{appB14}
\end{align}
so that Eqs.~\eqref{appB1}--\eqref{appB3} appear  unitless.

From Eqs.~\eqref{appB13} and \eqref{appB14} we obtain a connection between the grid spacing, particle mass, and charge.
We note first that the current density is
\be
\vec{J} = J_0 \Jhat = \hat{q} \frac{q_0}{\Delta x^3} \vhat \frac{\Delta x}{\Delta t} = \hat{q}\vhat \frac{q_0}{\Delta x^2 \Delta t},
\ee
i.e., $J_0 = q_0/\Delta x^2 \Delta t$. 
This implies that
\be
\frac{q_0}{\Delta x^2} = \frac{ m_0 c}{4\pi q_0 \Cfl \Delta t}
,\ee
i.e.,
\be
\Delta x = 4\pi \frac{q_0^2}{m_0} \left( \frac{\Cfl}{c} \right)^2.
\ee
Assuming that one computational macroparticle contains $N$ electrons (or positrons) we can write
\be\label{app:dx}
\Delta x = 4 \pi N \frac{q_\mathrm{e}^2}{m_\mathrm{e}} \left( \frac{\Cfl}{c} \right)^2
         = 4 \pi N \Cfl^2 r_\mathrm{e},
\ee
where we have set $m_0 = m_\mathrm{e}$ ($m_\mathrm{e}$ is the electron rest mass) and $q_0 = e$ ($e$ is the electron charge), and where $r = e^2/m_\mathrm{e} c^2 \approx \Ten{2.82}{-13}\cm$ is the classical electron radius.
This enables us to express the field scaling, Eq.~\eqref{appB13}, in physical (Gaussian) units 
\be
B_0 = \frac{m_\mathrm{e} c}{e \Cfl} \left(\frac{c}{\Cfl \Delta x} \right) = \frac{e }{r_\mathrm{e} \Cfl^2 \Delta x}.
\ee
The conversions from code units to Gaussian units can then be performed by selecting a length scale, $\Delta x$, as
\begin{align}
    \vec{B} &= B_0 \Bhat \approx \Ten{1.705}{3} \frac{\Bhat}{\Cfl^2}  \frac{1\cm}{\Delta x} \Gauss, \\
    \vec{E} &= B_0 \Ehat \approx \Ten{1.705}{3} \frac{\Ehat}{\Cfl^2}  \frac{1\cm}{\Delta x} ~\mathrm{statvolt}\,\cm^{-1}, \\
    \vec{J} &= \frac{B_0 }{4\pi \Delta t} \Jhat = \frac{e c}{4\pi r_\mathrm{e} \Cfl^3 \Delta x^2} \Jhat \approx \Ten{4.056}{12} \frac{\Jhat}{\Cfl^3} \left( \frac{1\cm}{\Delta x} \right)^2 ~\mathrm{statcoul}\,\mathrm{s}^{-1}, \\
    q &= \hat{q} \frac{e \Delta x}{4\pi \Cfl^2 r_\mathrm{e}} \approx \Ten{1.356}{2} \frac{\hat{q}}{\Cfl^2} \frac{\Delta x}{1\cm} ~\mathrm{statcoul}.
\end{align}

We next discuss some derived plasma quantities in the simulation.
The total relativistic plasma frequency is given as
\be\label{app:omegap_real}
\omega_\mathrm{p}^2 = \omega_{\mathrm{p},-}^2 + \omega_{\mathrm{p},+}^2 = 
\frac{4 \pi q^2 n}{\gammam m_\mathrm{e}} \left(1+\frac{m_-}{m_+}\right),
\ee
where $\omega_{\mathrm{p},-}$ and $\omega_{\mathrm{p},+}$ are the plasma frequencies of electrons and ions (or electrons and positrons), respectively, and where we have assumed $n = n_- = n_+$.
For electron-ion plasma $m_- = 1$ and $m_+ = 1836$ (or some smaller reduced proton mass), whereas for electron-positron pair plasma $m_-=1$, $m_+=1$, and $1 + m_-/m_+ = 2$.
Time dilatation from a relativistic bulk-motion (if any) is corrected by the mean Lorentz factor of that flow $\gammam$.
The counter-part in the code units is
\be\label{app:omegap}
\hat{\omega}_p^2 = \omega_p^2 \Delta t^2
= \frac{\hat{q}^2 N_{ppc}}{\gammam \hat{m}} \left(1+\frac{m_-}{m_+}\right)
= \frac{|\hat{q}| N_{ppc}}{\gammam} \left(1+\frac{m_-}{m_+}\right),
\ee
where $N_{ppc}$ is the number of particles per cell per species (again assuming charge equilibrium, $N_{ppc} = n_- = n_+$) and in the last step we  take into account that $|\hat{q}|/\hat{m} = 1$.
Initial numerical plasma oscillation frequency is obtained by specifying the skin-depth resolution, $\lambda_\mathrm{p} = c/\omega_\mathrm{p} = \mathcal{R}_\mathrm{p} \Delta x$, and remembering that $c = \Cfl \Delta x / \Delta t$ and $\omega_\mathrm{p} = \hat{\omega_\mathrm{p}} / \Delta t$, so that we obtain
\be
\hat{\omega}_{\mathrm{p},0} = \frac{\Cfl \Delta x}{\mathcal{R}_\mathrm{p} \Delta x} = \frac{\Cfl }{\mathcal{R}_\mathrm{p}}.
\ee
We can then fix the value of charge $|\hat{q}|$ (and $\hat{m}$) such that the initial plasma frequency for the electrons in the beginning of the simulation is (i.e., requiring $\Delta t = \hat{\omega}_{\mathrm{p},0}^{-1})$
\be
|\hat{q}| = \frac{\hat{\omega}_{\mathrm{p},0}^2 \gammam}{N_{ppc} \left(1+\frac{m_-}{m_+}\right)}.
\ee
Similarly, the relativistic species-specific cyclotron frequency is 
\be\label{app:omegaB_real}
\omega_B = \frac{q B}{m c \gamma},
\ee
which in code units is simply
\be\label{app:omegaB}
\omega_B \Delta t = \frac{\hat{q} \hat{B}}{\hat{m} \Cfl \gamma} \frac{m_-}{m_+},
\ee
where $\hat{m} \frac{m_+}{m_-}$ is the numerical mass of the particle (i.e., $1\hat{m}$ for pair plasma or $1836 \hat{m}$ for protons).
We note that the critical electron cyclotron frequency for $B = B_0$ (i.e., $\Bhat = 1$) corresponds to $\omega_{B,0} \Delta t = 1/\Cfl$.
Both Eqs.~\eqref{app:omegap} and \eqref{app:omegaB} can be translated to physical units by simply inserting numerical values to the time step $\Delta t^{-1} = c/\Cfl \Delta x$.
By comparing Eqs.~\eqref{app:omegap_real} vs. \eqref{app:omegap} and Eqs.~\eqref{app:omegaB_real} vs. \eqref{app:omegaB} we also see that numerical quantities are conveniently obtained by just replacing the physical quantities with the code variables.

Finally, the total plasma magnetization is given as
\be
\sigma = \frac{B^2}{4\pi n m_\mathrm{e} c^2 \gammam} = \left(\frac{\omega_B}{\omega_\mathrm{p}} \right)^2
 = \frac{\hat{B}^2}{N_{ppc} \hat{m} \Cfl^2 \gammam } \left( 1 + \frac{m_-}{m_+} \right)^{-1}.
\ee
This demonstrates the usefulness of the unit system.
Numerically we perform fewer floating-point operations because formulas appear unitless,
our typical values are of the same order of magnitude reducing floating-point round-off errors,
and physical quantities are easily transformed to the code unit system by replacing variables with code quantities (e.g., $x \rightarrow \hat{x}$ and $n=N_{ppc}$) and dropping $4\pi$ factors from the  Maxwell equations.

\section{Time and space staggering with finite-difference method}\label{app:fdtd}

Numerical charge conservation and divergence-free $\vec{B}$ field preservation follow from the use of a charge conserving current deposition strategy and the staggered Yee lattice \citep{yee1966}.
See \citet{esirkepov2001} for more in-depth discussion of the topic.

Let us define discrete versions of the $\nabla$ operator as 
\be
\nabla^+ A_{i,j,k} = \left( 
    \frac{A_{i+1,j,k} - A_{i,j,k}}{\Delta x}, 
    \frac{A_{i,j+1,k} - A_{i,j,k}}{\Delta y}, 
    \frac{A_{i,j,k+1} - A_{i,j,k}}{\Delta z}
                     \right)
\ee
and
\be
\nabla^- A_{i,j,k} = \left( 
    \frac{A_{i,j,k} - A_{i-1,j,k}}{\Delta x}, 
    \frac{A_{i,j,k} - A_{i,j-1,k}}{\Delta y}, 
    \frac{A_{i,j,k} - A_{i,j,k-1}}{\Delta z}
                     \right),
\ee
where $\Delta x$, $\Delta y$, and $\Delta z$ are the grid spacings.
These operators have the following properties
\be
\nabla^- \cdot \nabla^- \times = \nabla^+ \cdot \nabla^+ \times = 0 \quad \mathrm{and} \quad
\nabla^- \cdot \nabla^+ = \nabla^+ \cdot \nabla^-  = \hat{\Delta},
\ee
where $\hat{\Delta}$ is the discrete Poisson operator in central differences such that
\be
\hat{\Delta} A_{i,j,k} = 
\frac{A_{i-1, j,k} - 2 A_{i,j,k} + A_{i+1, j,k}}{\Delta x^2} +
\frac{A_{i,j-1,k} - 2 A_{i,j,k} + A_{i, j+1,k}}{\Delta y^2} +
\frac{A_{i,j,k-1} - 2 A_{i,j,k} + A_{i+1, j,k+1}}{\Delta z^2}.
\ee

Let us stagger our variables in space and time such that 
\begin{align}
    \vec{x} &= \vec{x}^n,  \\
    \vec{u} &= \vec{u}^{n+\hf}, \\
    \rho &= \rho_{i+\hf, j+\hf, k+\hf}^n,  \\
    \vec{J} &= (J_{x;~ i, j+\hf, k+\hf}, J_{y;~ i+\hf, j, k+\hf}, J_{z;~ i+\hf, j+\hf, k})^{n+\hf}, \\
    \vec{E} &= (E_{x;~ i, j+\hf, k+\hf}, E_{y;~ i+\hf, j, k+\hf}, E_{z;~ i+\hf, j+\hf, k})^{n},  \\
    \vec{B} &= (B_{x;~ i+\hf, j, k}, B_{y;~ i, j+\hf, k}, B_{z;~ i, j, k+\hf})^{n+\hf}.
\end{align}
Maxwell's equations for $\vec{E}$ and $\vec{B}$ are given as
\be\label{eq:fdE2}
\frac{\vec{E}^{n+1} - \vec{E}^n}{\Delta t} = \nabla^+ \times \vec{B}^{n+\hf} - 4\pi\vec{J}^{n+\hf},
\ee
\be\label{eq:fdB2}
\frac{\vec{B}^{n+\hf} - \vec{B}^{n-\hf}}{\Delta t} = -\nabla^- \times \vec{E}^{n},
\ee
\be\label{eq:divEfdtd}
\nabla^+ \cdot \vec{E}^n = 4\pi \rho^n,
\ee
\be
\nabla^- \cdot \vec{B}^{n+\hf} = 0.
\ee

The discrete form of the charge conservation then follows by operating with $\nabla^+ \cdot$ on Eq.~\eqref{eq:fdE2} and substituting \eqref{eq:divEfdtd} into the equation that follows. 
We then obtain
\be
\frac{\rho^{n+1} - \rho^n}{\Delta t} + \nabla^+ \cdot \vec{J}^{n+\hf} = 0.
\ee Similarly, operating on Eq.~\eqref{eq:fdB2} with $\nabla^-$ we obtain the divergence-free nature of $\vec{B}$ field
\be
\frac{\nabla^- \vec{B}^{n+\hf} - \nabla^- \vec{B}^{n-\hf}}{\Delta t} = 0.
\ee
Thus, if 
\be
\nabla^+ \cdot \vec{E} = 4\pi \rho
\ee
and
\be
\nabla^- \cdot \vec{B} = 0
\ee
at $t=0$, then the divergence of $\vec{E}$ is always equal to the charge density and $\vec{B}$ will retain its divergence-free nature up to machine precision during the simulation.

\section{Relativistic Boris scheme}\label{app:Boris}
We  present  the full relativistic Boris scheme here that updates particle four-velocity one time step forward.
This can be found as \textsc{Pusher} solver named \texttt{BorisPusher} in the \textsc{pic} module.
In the beginning we have a particle with a four-velocity $\vec{u}^n$ in units of $c$.
The Lorentz update is done via velocities in units of $\Cfl$ since this is the effective speed of light experienced by the electromagnetic fields.
To differentiate between these two units we use primed quantities to mark velocities in units of $\Cfl$.

The scheme begins with a first half of electric field acceleration as
\be\label{eq:appB1}
\vec{u}_1' = \Cfl \vec{u}_0 + \frac{1}{2} \hat{s} \Ehat,
\ee
where 
\be
\hat{s} = \frac{ \hat{q} }{\hat{m}_s m_{\pm}} = \frac{\mathrm{sign}\{ \hat{q}_s \} }{m_{\pm}},
\ee
and $m_{\pm} = m/m_\mathrm{e}$ is the mass of the particle in units of electron rest mass $m_\mathrm{e}$.
This four-velocity in units of $\Cfl$ corresponds to a Lorentz factor of
\be
\gamma_1 = \frac{\sqrt{\Cfl^2 + u_1'^2}}{\Cfl} = \sqrt{1 + u_1'^2/\Cfl^2}.
\ee
We then proceed with a first half of magnetic rotation 
\be
\vec{u}_2' = \vec{u}_1' f + \hat{s} \frac{\vec{u}_1'}{\Cfl \gamma} \times \frac{1}{2} \Bhat f,
\ee
where 
\be
f = \frac{2}{1 + \left( \frac{\Bhat}{2\Cfl\gamma_1} \right)^2 }.
\ee
As a final step we combine the second half of magnetic rotation and second half of the electric field acceleration as
\be
\vec{u}_3' = \vec{u}_1' + \hat{s} \frac{\vec{u}_2'}{\Cfl \gamma} \times \frac{1}{2} \Bhat + \frac{1}{2} \hat{s} \Ehat.
\ee
Final particle four-velocity in units of $c$ is then 
\be\label{eq:appB2}
\vec{u}^{n+1} = \frac{u_3'}{\Cfl},
\ee
summarizing a recipe for updating the particle velocity from an initial time $\vec{u}^n$ to $\vec{u}^{n+1}$ (i.e., one time step of $\Delta t$ forward).

\section{Current deposition}\label{app:zigzag}

We  summarize here the charge-conserving ZigZag current deposition scheme \citep{Umeda_2003} implemented as the  \texttt{ZigZag} solver in \runko.
First, we compute the relay point $\vec{x}_\mathrm{r}$ between previous particle position $\vec{x}_1$ (and closest grid index $\vec{i}_1 = (i_1,j_1,k_1)$) and the current position $\vec{x}_2$ (and closest grid index $\vec{i}_2$) as
\be\label{eq:appE1}
\vec{x}_\mathrm{r} = \mathrm{min} \left\{ 
    \mathrm{min}\left\{ \vec{i}_1,\vec{i}_2 \right\}+1, \mathrm{max}\left\{ \mathrm{max}\left\{ \vec{i}_1,\vec{i}_2 \right\},\frac{1}{2}\left(\vec{x}_1+\vec{x}_2 \right) \right\} \right\}.
\ee
Here we assume element-wise $\mathrm{min}$ and $\mathrm{max}$ operators as $\mathrm{min}\{ \vec{a}, \vec{b}\} \equiv (\mathrm{min}\{ a_x, b_x \},  \mathrm{min}\{ a_y, b_y \},  \mathrm{min}\{ a_z, b_z \})$.

The current fluxes are then obtained as
\begin{align}
    \vec{F}_1 &= \hat{q}(\vec{x}_\mathrm{r} - \vec{x}_1), \\
    \vec{F}_2 &= \hat{q}(\vec{x}_2 - \vec{x}_\mathrm{r}).
\end{align}
We assume first-order shape functions for the particles.
This is related to particle weights defined at the mid points $(\vec{x}_\mathrm{r} + \vec{x}_1)/2$ and $(\vec{x}_2 + \vec{x}_r)/2$ as
\begin{align}
    \vec{W}_1 &= \frac{1}{2} \left( \vec{x}_1 + \vec{x}_\mathrm{r} \right) - \vec{i}_1, \\
    \vec{W}_2 &= \frac{1}{2} \left( \vec{x}_2 + \vec{x}_\mathrm{r} \right) - \vec{i}_2. \label{eq:appE2}
\end{align}
Depending on the dimensionality, the weights are described as $\vec{W}_{1,2} = (W_{1,2;x})$ for $D=1$,  $\vec{W}_{1,2} = (W_{1,2;x}, W_{1,2;y})$ for $D=2$, or $\vec{W}_{1,2} = (W_{1,2;x}, W_{1,2;y}, W_{1,2;z})$ for $D=3$.
Finally, the current density $J = F W$ needs to be injected into adjacent grid points following a charge-conserving deposition strategy.
This can be implemented following \citet{Umeda_2003}.

\end{appendix}

\end{document}